\documentclass{hobbs}

\usepackage{graphicx,times}
\usepackage[numbers,super]{natbib}
\usepackage{CJK,color}
\usepackage{url}
\usepackage{fancyhdr}
\usepackage[T1]{fontenc}
\usepackage{fancyvrb}
\usepackage{verbatim}

\providecommand{\bm}[1]{\mbox{\boldmath$#1$\unboldmath}}

\begin{document}
\makeatletter
\def\@cite#1#2{\textsuperscript{[{#1\if@tempswa  #2\fi}]}}
\makeatother
\begin{CJK*}{GBK}{song}

   \title{Using the Parkes Pulsar Data Archive
}


   \volnopage{Vol. 9 (2012) No.3, 229--235}
   \setcounter{page}{229}          
   \author{
  J. Khoo
    \inst{1},
   G. Hobbs
    \inst{1},
  R. N. Manchester
    \inst{1},
  D. Miller
    \inst{2},
  J. Dempsey
    \inst{2}
   }
   \institute{CSIRO Astronomy and Space Science, Australia Telescope National Facility, P.O. Box 76, Epping, NSW 1710, Australia
   \and CSIRO Information Management \& Technology, P.O. Box 225, Dickson ACT 2602, Australia
   }
   \date{Received~~2012 month day; accepted~~2012~~month day}    
   \abstract{ The Parkes Pulsar Data Archive currently provides access to 165,755 data files obtained from observations carried out at the Parkes Observatory since the year 1991. Data files and access methods are compliant with the Virtual Observatory protocol. This paper provides a tutorial on how to make use of the Parkes Pulsar Data Archive and provides example queries using on-line interfaces.
   }


   \maketitle
\end{CJK*}
%
%

\section{Introduction}\label{sec:intro}

The Parkes Pulsar Data Archive provides public access to 165,755 data files related to observations of pulsars recorded at the Parkes observatory. It is part of the Data Access Portal---an on-line interface (http://data.csiro.au) that provides access to published Commonweath Scientific and Industrial Research Organisation (CSIRO) data sets over a range of science disciplines. The Parkes Pulsar Data Archive is the largest publicly accessible database of pulsar data files. Approximately $ 10^{5} $ data files contain observations from surveys of the sky and the remainder are for observations of 775 known pulsars and their corresponding calibration signals.

Improvements to the system have been made since the initial description of the archive in Hobbs et al. (2011)\cite{Hobbs et al.(2011)}. This paper focuses on how to use the current Parkes Pulsar Data Archive and provides walk-throughs of common use cases. \S~\ref{sec:contents} describes the types of data that can be obtained from the data archive. \S~\ref{sec:nav-pulsar-search} explains how to navigate to the Pulsar Search from the main Data Access Portal page. \S~\ref{sec:examples} shows examples of how to obtain data used in complemetary data-processing tutorials \cite{k12,h12} from the on-line interface. \S~\ref{sec:vo} provides examples of how to use various Virtual Observatory tools to access contents of the data archive.

\begin{figure}
\centering\includegraphics[width=150mm]{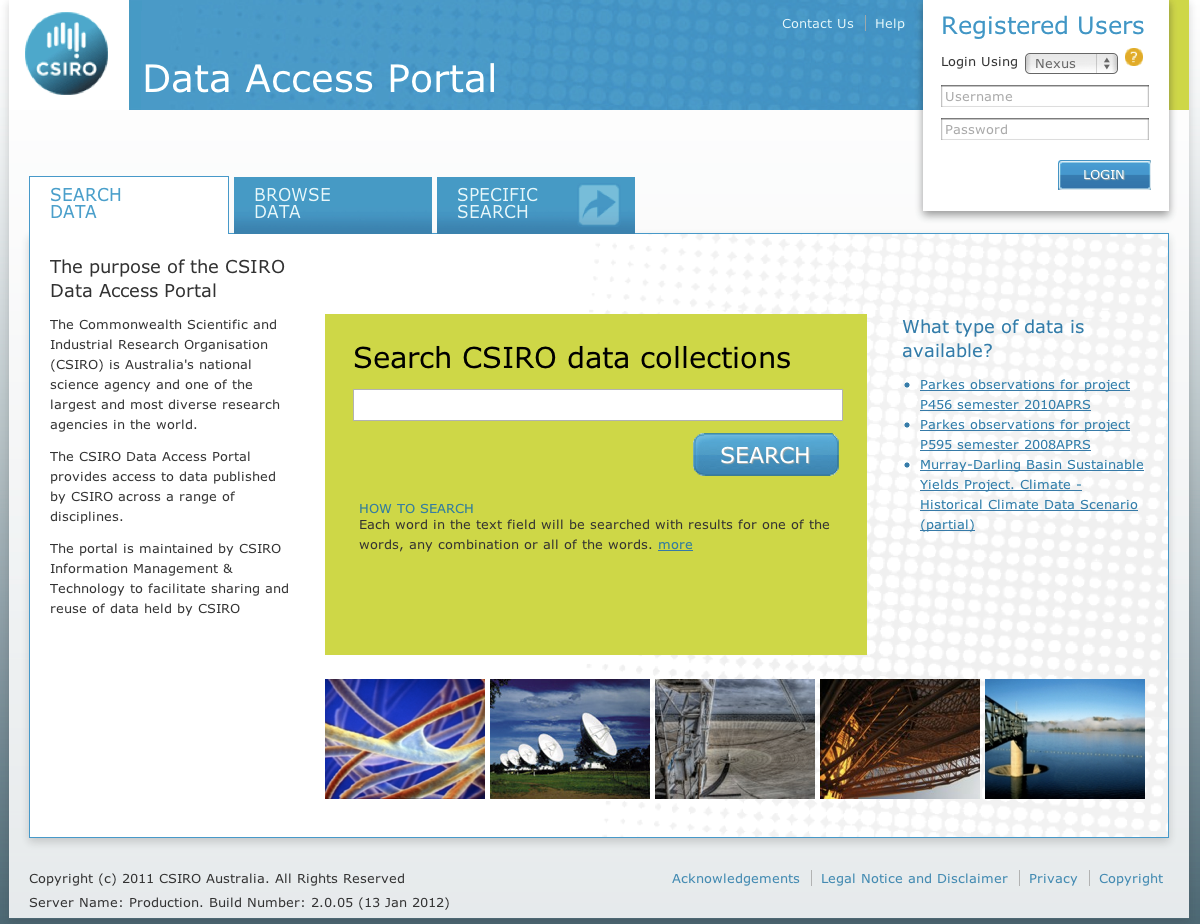}
\caption{CSIRO's Data Access Portal (http://data.csiro.au)}
\label{fig:dap}
\end{figure}

\begin{figure}
\centering\includegraphics[width=150mm]{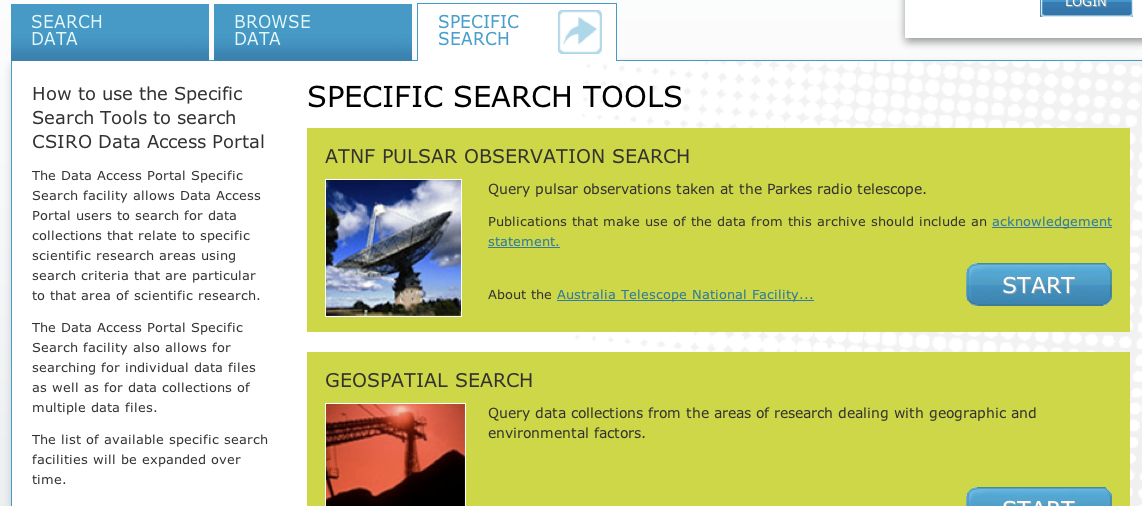}
\caption{Data Access Portal: Specific Search}
\label{fig:spec-search}
\end{figure}

\begin{figure}
\centering\includegraphics[width=130mm]{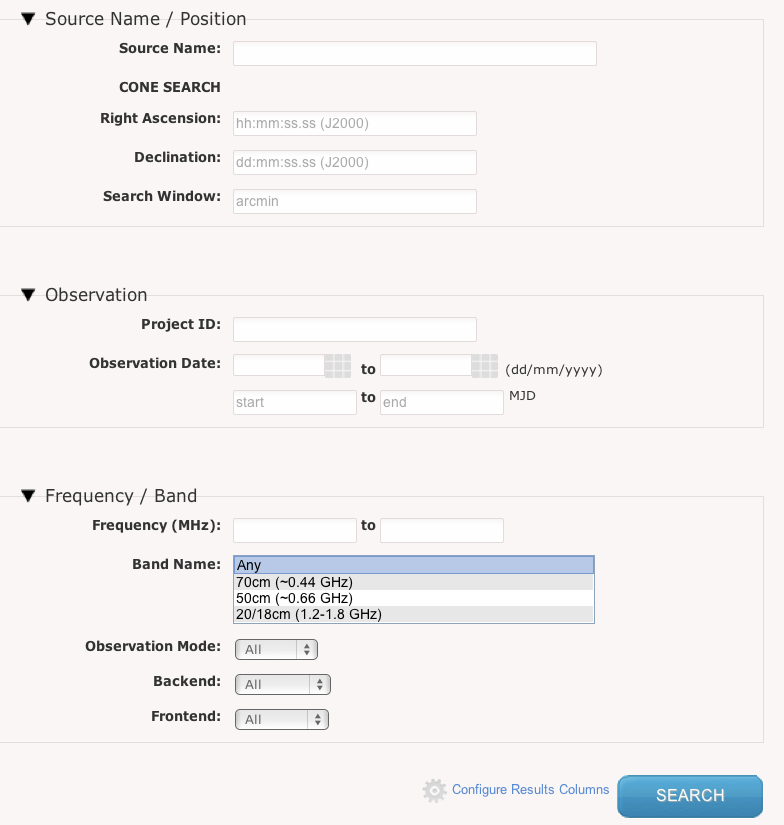}
\caption{Pulsar Search: main window}
\label{fig:pulsar-search-main}
\end{figure}

\begin{figure}
\centering\includegraphics[width=110mm]{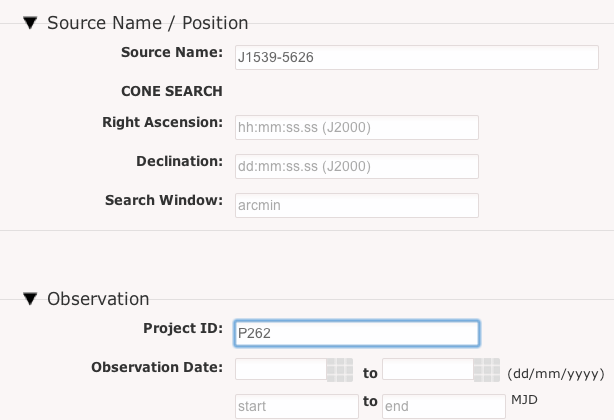}
\caption{Search criteria for Hobbs et al. 2012}
\label{fig:hobbs-main}
\end{figure}

\begin{figure}
\centering\includegraphics[width=150mm]{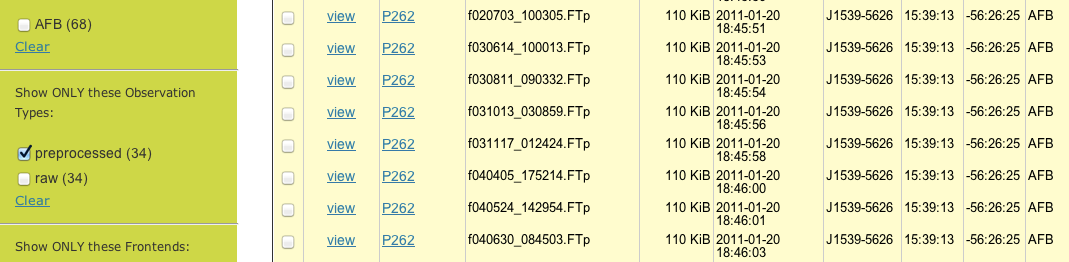}
\caption{Facet search options for Hobbs et al. 2012}
\label{fig:hobbs-facet}
\end{figure}

\begin{figure}
\centering\includegraphics[width=150mm]{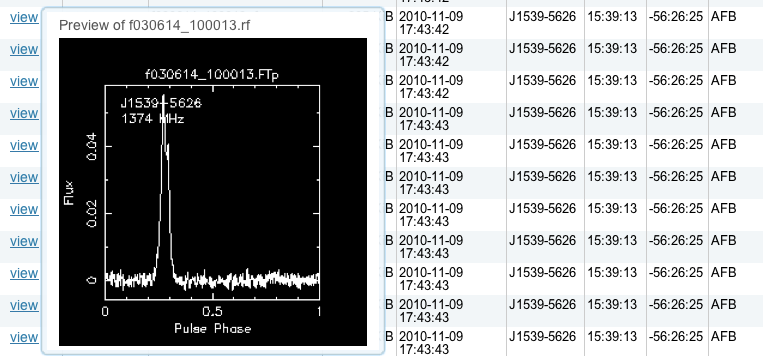}
\caption{Thumbnail preview of PSR J1539$-$5628 for Hobbs et al. 2012}
\label{fig:hobbs-preview}
\end{figure}

\section{Archive Contents}\label{sec:contents}

There are two main observing modes that are used in pulsar astronomy: 1) ``fold mode'' that results in a single pulse profile for each observation; such files are used for calculating pulse arrival times and studies utilising pulse profiles, and 2) ``search mode'' a set of time series recording the signal from the telescope using a selected number of bits, sample interval, and frequency resolution.  This mode is used, primarily, for discovering new pulsars and studying single pulses from pulsars. The data formats for both modes are based on the PSRFITS definition \cite{Hotan et al.(2004)}. In addition to the raw observations of fold-mode data, the Parkes Pulsar Data Archive also contains the corresponding compressed representation of the data; the compression performed is integration over the polarsation, time, and frequency domains. This gives the user the option to opt for downloading the compressed file---decreasing the download time---rather than the raw data.



Pulsar data recorded at Parkes are automatically added to the Parkes Pulsar Data Archive; however, the archive complies to the 18-month embargo policy enforced by CSIRO such that only members of the observing project have access to observations recorded in the last 18 months. An exemption to this rule is data recorded by the observing project PULSE@Parkes \cite{Hobbs et al.(2009)}; since the project is for science outreach and the education of high-school students and teachers, the data are immediately available.

\section{Navigating to the Pulsar Search}\label{sec:nav-pulsar-search}

Follow the instructions below to access the main pulsar search page:

\begin{enumerate}
\item Navigate a web browser to http://data.csiro.au.
\item From the Data Access Portal home page (Figure \ref{fig:dap}), click `SPECIFIC SEARCH' to access the various data access points.

\item From the `SPECIFIC SEARCH TOOLS' page (Figure \ref{fig:spec-search}), click `ATNF PULSAR OBSERVATION SEARCH' to access the Pulsar Search window (Figure \ref{fig:pulsar-search-main}).

\end{enumerate}

\section{Examples}\label{sec:examples}

This section will provide two step-by-step procedures to obtain data.

\subsection{Hobbs et al. (2012): PSR J1539$-$5626 Preprocessed Fold-mode Data}\label{sec:example}

The data used in Hobbs et al. (these proceedings) are preprocessed (with the data integrated in time, frequency, and polarisation) observations of PSR J1539$-$5626. These observations are part of the P262 project: ``Timing of young pulsars''. Data files can be obtained via the following instructions:

\begin{enumerate}
\item Proceed to the Pulsar Search webpage using a web browser.
\item Enter the following search criteria in the main pulsar search window: J1539$-$5626 for source name, P262 for project ID (as shown in Figure~\ref{fig:hobbs-main}), and then click ``Search''.

\item New tables will appear on the screen showing a summary of query results. On the left-hand table, untick ``raw'' so that only preprocessed files are displayed (shown in Figure~\ref{fig:hobbs-facet}). This will refine the search and remove all raw data files (leaving only preprocessed observations of PSR J1539$-$5626 from project P262).

\item Press the icon under the ``Preview'' column to see a preview of the pulse profile. Select the desired files to download by ticking the toggle box on the left-hand side (shown in Figure~\ref{fig:hobbs-preview}). Click ``Download'' to  copy the selected files onto your computer.

\end{enumerate}

Once fold-mode data have been obtained, PSRCHIVE \cite{Hotan et al.(2004)} can be used to calculate pulsar period variations. Once pulse arrival times have been calculated, TEMPO2 \cite{Hobbs et al.(2006)} can be used to study the pulse arrival times.

\begin{figure}
\centering\includegraphics[width=130mm]{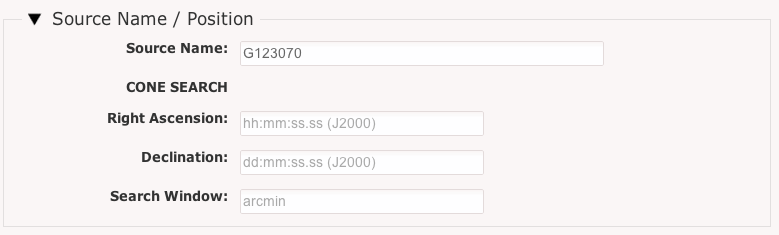}
\caption{Search critera for Keith et al. (2012)}
\label{fig:keith-main}
\end{figure}

\begin{figure}
\centering\includegraphics[width=150mm]{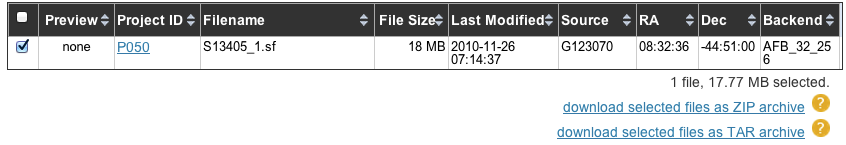}
\caption{Query results for Keith et al. (2012)}
\label{fig:keith-results}
\end{figure}

\subsection{Keith et al. (2012): PSR J0835$-$4510 Search-mode Data}\label{sec:example}

The data file used in Keith et al. (2012) is a search-mode observation near the Vela Pulsar (PSR J0835$-$4510). These observations were recorded in the P050 project: ``Parkes Southern Pulsar Survey''. It can be obtained via the following instructions

\begin{enumerate}
\item Proceed to the Pulsar Search webpage using a web browser (see \S~\ref{sec:nav-pulsar-search}).
\item Enter the following search criteria in the main pulsar search window: G123070 for source name, P050 for project ID (as shown in Figure~\ref{fig:keith-main}), and then click ``Search''.
\item Select the first result by ticking the checkbox on the left-hand side of the table (as shown in Figure~\ref{fig:keith-results}). Scroll down to the bottom of the page and click ``Download'' to copy the files onto your computer.
\end{enumerate}

Once search-mode data have been obtained, SIGPROC \cite{Lorimer(2011)} or PRESTO \cite{Ransom(2011)} can be used to process the data and search for pulsars.

\section{Virtual Observatory}\label{sec:vo}
\subsection{Introduction}\label{sec:vo-intro}

The Virtual Observatory (VO) is a set of astronomy-related data archives and software tools. VO software tools have the capability to perform complex operations on any combination of the accessible VO data archives. The contents of a VO data archive adhere to a defined set of protocols; this allows development of VO tools to follow the VO definition and, as a result, perform operations (such as statistical analysis, plotting, etc.) on contents of VO databases.

This section provides the steps required to access the Parkes Pulsar Data Archive using two VO tools from the AstroGrid software package (http://www.astrogrid.org): VODestop and TopCat. VODesktop is used as the medium to communicate with the Parkes Pulsar Data Archive. You will be shown how to perform a example query to obtain a VOTable (a VO-formatted file representing the results of the query). We will then use TopCat to analyse and plot the results.

\subsection{VODesktop: Performing a query}\label{sec:vo-table}

\begin{figure}
\centering\includegraphics[width=155mm]{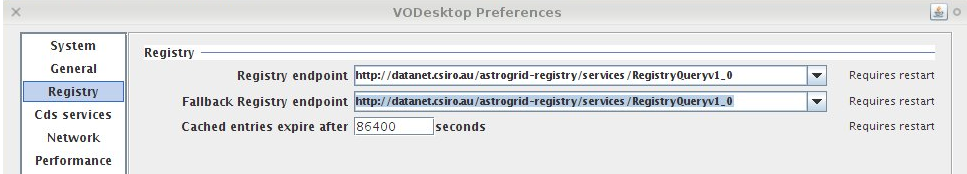}
\caption{Endpoint registry prefences for VODesktop.}
\label{fig:vodesktop-pref}
\end{figure}

\begin{figure}
\centering\includegraphics[width=120mm]{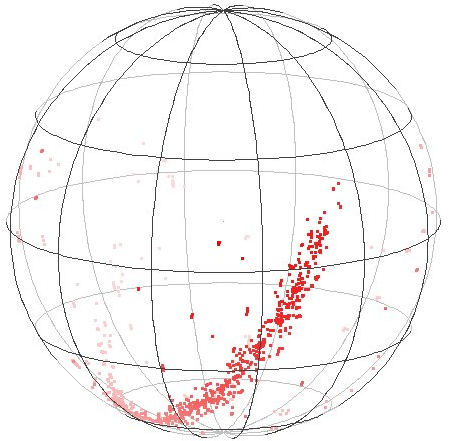}
\caption{Positions of the data in the Parkes Pulsar Data Archive.}
\label{fig:topcat-sphere}
\end{figure}

\begin{figure}
\centering\includegraphics[width=110mm]{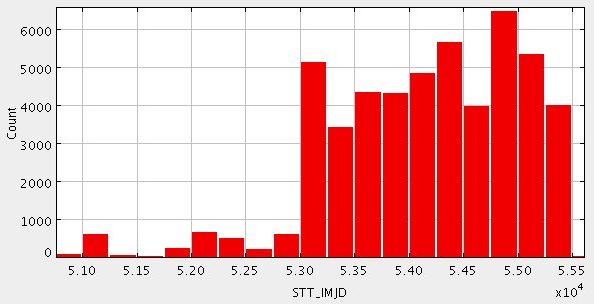}
\caption{Distribution of number of observations versus time.}
\label{fig:topcat-histogram}
\end{figure}

\begin{enumerate}
\item Go to http://www.astrogrid.org/wiki/Install/Downloads and download VODesktop.
\item Open VODesktop and go to Preferences -$>$ Registry.
\item Change the both registry endpoints to http://datanet.csiro.au/astrogrid-registry/services/RegistryQueryv1\_0 (as shown in Figure~\ref{fig:vodesktop-pref}).
\item Click Apply and restart VODesktop.
\item Select Recent Changes and click on ANDS Parkes Data Archive.
\item Under Actions, click ADQL Query.
\item In the Task Runner window, change the query text to:
SELECT * FROM observation
\item Choose to save the results, locally, as a VOTable.
\end{enumerate}

\subsection{TopCat: Analysing the results}\label{sec:vo-topcat}

\begin{enumerate}
\item Go to http://www.astrogrid.org/wiki/Install/Downloads and download TopCat.
\item To load the saved VOTable, File -$>$ Load Table -$>$ Filestore Browser and navigate to where you saved the VOTable.
\end{enumerate}

Once the VOTable has been read by TopCat, TopCat's plotting functionality become available. Figure~\ref{fig:topcat-sphere} shows the pulsar positions on the 3D galactic coordinate sphere using the Spherical Plot. Figure~\ref{fig:topcat-histogram} shows a histogram of all pulsar observations and observation date.

\section{Future Development}\label{sec:future}

The Parkes Pulsar Data Archive will undergo expansion to include timing data from ATNF Parkes Swinburne Recorder and CASPER Parkes Swinburne Recorder; and (potentially) the Berkeley Parkes Swinburne Recorder recorded at the Parkes 64-m telescope. Further conversions between data formats will also be performed to include historical data from earlier projects. A related database of pulsar profiles is being developed at CSIRO Astronomy and Space Science (CASS). This tool will allow users to search pulse profiles of known pulsars and see visual representations when comparing parameters between pulsars. CASS is also developing a data access portal that provides access to various non-pulsar radio-astronomy files of interest to external collaborators of CASS staff members. These systems will use the Parkes Pulsar Data Archive as a guide and will, also, be compatible with the Virtual Observatory formats and protocols.

\section{Conclusion}\label{sec:conclusion}

The Parkes Pulsar Data Archive currently hosts 165,755 data files---search mode and fold mode. These data are publicly accessible. The data archive plays an important role in providing researchers worldwide with observations recorded at Parkes Observatory. The Parkes Pulsar Data Archive is an effective solution for our current data sets (pulsar observations from Parkes Observatory since 1991). With ASKAP and SKA being built, a different set of challenges stand ahead to produce similar software solutions.

\normalem
\begin{acknowledgements}

This project is supported by the Australian National Data Service (ANDS). ANDS is supported by the Australian Government through the National Collaborative Research Infrastructure Strategy Program and the Education Investment Fund (EIF) Super Science Initiative19.  We acknowledge the software development provided by the CSIRO IM\&T Software Services, the business process development by the CSIRO IM\&T Data Management Service and project management through Citadel Systems. This research has made use of software provided by the UK's AstroGrid Virtual Observatory Project, which is funded by the Science and Technology Facilities Council and through the EU's Framework 6 programme. The Parkes radio telescope is part of the Australia Telescope, which is funded by the Commonwealth of Australia for operation as a National Facility managed by the Commonwealth Scientific and Industrial Research Organisation (CSIRO). GH is the recipient of an Australian Research Council QEII Fellowship (\#DP0878388).

\end{acknowledgements}

\bibliographystyle{nature,unsrt}

\label{lastpage}

\end{document}